\documentclass{ieeeaccess}
\usepackage{cite}
\usepackage{amsmath,amssymb,amsfonts}
\usepackage{algorithmic}
\usepackage{graphicx}
\usepackage{textcomp}
\usepackage{blindtext}
\usepackage[T1]{fontenc}
\usepackage{epstopdf}
\usepackage{multirow}
\usepackage{booktabs}
\usepackage{ragged2e}
\usepackage[inline]{enumitem}
\usepackage{diagbox}
\usepackage{xcolor}

\newcommand\blfootnote[1]{%
	\begingroup
	\renewcommand\thefootnote{}\footnote{#1}%
	\addtocounter{footnote}{-1}%
	\endgroup
}

\newcommand{\change}[1]{#1}

\def\BibTeX{{\rm B\kern-.05em{\sc i\kern-.025em b}\kern-.08em
		T\kern-.1667em\lower.7ex\hbox{E}\kern-.125emX}}
\begin{document}
	\history{Date of first submission: Nov. 22, 2018; date of resubmission:  Jan. 17, 2019; date of acceptance: Feb. 02, 2019}
	\doi{}
	
	\title{Context-Awareness Enhances 5G Multi-Access Edge Computing Reliability}
	\author{\uppercase{Bin Han}\authorrefmark{1}\IEEEmembership{Member, IEEE}, \uppercase{Stan Wong}\authorrefmark{2}, \uppercase{Christian Mannweiler}\authorrefmark{3},\\
	\uppercase{Marcos Rates Crippa}\authorrefmark{1},	\uppercase{and Hans D. Schotten}\authorrefmark{1,4}\IEEEmembership{Member, IEEE}}
	\address[1]{Institute of Wireless Communication, Technische Universit\"at Kaiserslautern, 67655 Kaiserslautern, Germany}
	\address[2]{UbiXpace, E3 4PU London, UK.}
	\address[3]{Nokia Bell Labs Germany, 81541 Munich, Germany}
	\address[4]{Research Group Intelligent Networks, German Research Center for Artificial Intelligence (DFKI GmbH), 67663 Kaiserslautern, Germany.}
	\tfootnote{\justify This work has been performed in part in the framework of the H2020-ICT-2014-2 project 5G NORMA [grant number 671584]. The authors would like to acknowledge the contributions of their colleagues. This information reflects the consortium's view, but the consortium is not liable for any use that may be made of any of the information contained therein.}
	
	\markboth
	{B. Han \headeretal: Context-Awareness Enhances 5G MEC Reliability}	{}
	
	\corresp{Corresponding author: Bin Han (e-mail: binhan@eit.uni-kl.de).}
	
	\begin{abstract}
	The fifth generation (5G) mobile telecommunication network is expected to support \change{Multi-Access Edge Computing} (MEC), which intends to distribute computation tasks and services from the central cloud to the edge clouds. Towards ultra-responsive, ultra-reliable and ultra-low-latency MEC services, the current mobile network security architecture should enable a more decentralized approach for authentication and authorization processes. This paper proposes a novel \change{decentralized} authentication architecture that supports flexible and low-cost local authentication with the awareness \change{of} context information of network elements such as user equipment and virtual network functions. \change{Based on a Markov model for backhaul link quality, as well as a random walk mobility model with mixed mobility classes and traffic scenarios, numerical simulations have demonstrated that the proposed approach is able to achieve a flexible balance between the network operating cost and the MEC reliability.}
	\end{abstract}
	
	
	\titlepgskip=-15pt
	
	\maketitle
	
	\section{Introduction}
	\IEEEPARstart{F}{ifth} generation (5G) mobile telecommunication network has been expected to involve \change{Multi-Access Edge} Computing (MEC) \cite{barbarossa2014communicating,sabella2016mobile,corcoran2016mobile}. MEC has evolved from mobile cloud computing, which shifts the important computation and storage tasks from the core network to the access network. Thus, mobile services can gain \change{superiority} of low latency, power efficiency, context-awareness, and enhanced privacy protection~\cite{mao2017survey,abbas2018mobile}. Furthermore, the security concerns become across the core network and the access network -- from network elements to user equipment (UE) and from tenants to their network slices~\cite{lal2017nfv}. For example, the mobility management entity (MME) is set to be divided into few functions which can be co-existed in the access network and core network. This function decomposition gives the overall system flexibility, deployment efficiency and service agility.
	
	The fourth generation (4G) long-term evolution (LTE) authentication and key agreement (AKA) has an architecturally inherited dependency on the core network, and can therefore constrain the exploitation of MEC. Nevertheless, there has been limited contributions on the evolution of the LTE centralized security architecture despite of demands of improved AKA efficiency in MEC~\cite{li2018security}. However, delivering a flexible next generation telecommunication system is the main objective of 5G. All services at the edge of the network should be securely provided. Hence, the authentication and authorization processes can also exist at the edge of the network to increase the subscriber authentication efficiency and the reliability to the overall system.
	
	\blfootnote{This is a pre-print for private use only. \copyright IEEE 2019}
	
	Generally, authentication, authorization and accounting (AAA) processes require certain amount of cost from the system per registration of a subscriber. Putting these processes at the edge of the network would increase the efficiency of subscriber authentication and authorization. Also collecting valuable information helps to understand the context of the overall network situations, the connectivity availability and the traffic pattern~\cite{klein2010access}. With such knowledge we can increase the overall system reliability with minimal operations cost. In this paper, we will use the context-awareness technique to extract the network situation and reduce the traffic in between \change{fronthaul} and \change{backhaul} when subscriber authentication and authorization take place. Particularly, we use network context information to distinguish the normal from emergency situation and deliver proactive control in facing any situation.
	
	\change{The main contribution of this paper is a novel decentralized security approach for MEC in 5G networks, which aims at delivering a flexible and context-aware AKA executing within an edge cloud (EC), and decouples the authentication process from the central cloud when backhaul connection outages are detected. The propose approach is to make aware of network context information and able to proactively synchronize subscriber profiles to support local AKA procedures. We develop an numerical simulation demonstrates the balance in between the MEC reliability and the operations cost.}
	
	The paper is organized as follows. Section \ref{sec:state_of_the_art} briefly reviews the 4G LTE AKA mechanism, and analyzes its inadequacy of 5G MEC services. In Section \ref{sec:architecture}, we propose a novel decentralized authentication architecture in 5G. A Virtualized AAA system across the central cloud and edge cloud is elaborated in Section \ref{subsec:5gaaa}. Section \ref{subsec:tz} formulates a MEC cognitive security system that maintains the availability of the services. Section \ref{subsec:sync} presents a context-aware mechanism that enables the local authentication to reduce the backhaul traffic and improve the efficiency of subscriber management. Section \ref{sec:model} presents a Markov chain to characterize and forecast the reliability of backhaul connections. {At the end, Section \ref{sec:simulation} demonstrates the effectiveness of our proposed method with numerical simulations, before we} conclude this paper and provide a couple of future challenges in Section \ref{sec:conclusion}. \change{For the readers' convenience, Tab.~\ref{tab:acronyms} lists all the abbreviations used in this article in the alphabetical order.}
	\begin{table}[!htbp]\label{tab:acronyms}
		\begin{tabular}{ll}
			\toprule[1.5px]
			Abbreviation&Definition\\
			\midrule[1.2px]
			3GPP&Third Generation Partnership Project \\
			4G&fourth generation\\
			5G&fifth generation\\
			AAA&authentication, authorization and accounting\\
			AKA&authentication and key agreement\\
			AMF&access and mobility management function\\
			AS&access stratum\\
			BSS&business supporting system\\
			CCCM&Central Cloud Connection Monitoring\\
			CSSO&Central Security Service Outage\\
			CSSR&Central Security Service Risk\\
			EC&edge cloud\\
			ES&Emergency Service\\
			EPS&Evolved Packet System\\
			ETSI&European Telecommunications Standards Institute\\
			IKE&Internet key exchange \\
			IoT-GW&Internet-of-Things Gateway\\
			IPsec&Internet protocol security\\
			LAA&Local Access Assistant\\
			LSS&Local Subscriber Servers\\
			LTE&Long-Term Evolution\\
			MEC&Multi-Access Edge Computing\\
			MME&Mobility Management Entity\\
			MNO&mobile network operator\\
			NAS&non-access stratum\\
			NFV MANO&network function virtualization management and\\
			&orchestration\\
			OSS&operations support system\\
			O\&M&orchestration and management\\
			RAN&radio access network\\
			SA&Security Auditing\\
			SDM&software-defined mobile network\\
			SDMO&SDM orchestrator\\
			SDM-C&SDM coordinator\\
			SDM-X&SDM controller\\
			SDN&software-defined network\\
			SEG&security gateway\\
			SMS&short message service\\
			TZ&Trust Zone\\
			UDM&unified data management \\
			UE&user equipment\\
			VIM&virtual infrastructure manager\\
			VNF&virtual network function\\
			V-AAA&Virtualized-AAA\\
			ZM&Zone Management\\
			\bottomrule[1.5px]
		\end{tabular}
		\caption{List of acronyms}
	\end{table}
	
	\section{Security Architecture of Cellular Networks}\label{sec:state_of_the_art}
	\subsection{Authentication and Key Agreement in the 3GPP EPS}
	Security in the Third Generation Partnership Project (3GPP) Evolved Packet System (EPS) is divided into access stratum (AS) and non-access stratum (NAS) security. Security procedures rely on the EPS architecture; the most important elements are shown in Fig. \ref{fig:LTE-sec} and \cite{3gpp-33401,3gpp-33402}. 
	
	When a user equipment (UE) attaches to a LTE network, the AKA is executed between UE and Mobility Management Entity (MME). \change{Using the same so-called key of access security management entities ($K_{\text{ASME}}$), keys for securing the NAS signaling are derived. Using $K_{\text{ASME}}$, UE and MME also derive a key $K_{\mathrm{eNB}}$, which is also passed to the evolved Node B (eNB) that the UE is connecting to. $K_{\mathrm{eNB}}$} is used in the communication session as the basis of AS security, i.e. security on the radio link between the UE and the eNB. Three further keys are derived from $K_{\mathrm{eNB}}$ \change{for control plane integrity protection, control plane encryption, and user plane encryption, respectively. A fourth key for user plane integrity protection is derived from $K_{\mathrm{eNB}}$ only when required. These traffic protection keys are not used for an infinite amount of data. The eNB periodically initiates intra-cell handover to derive a new $K_{\mathrm{eNB}}$ and new protection and encryption keys. }
	
	\Figure[htbp!](topskip=0pt, botskip=0pt, midskip=0pt)[trim={0 2cm 0 0},width=.9\textwidth]{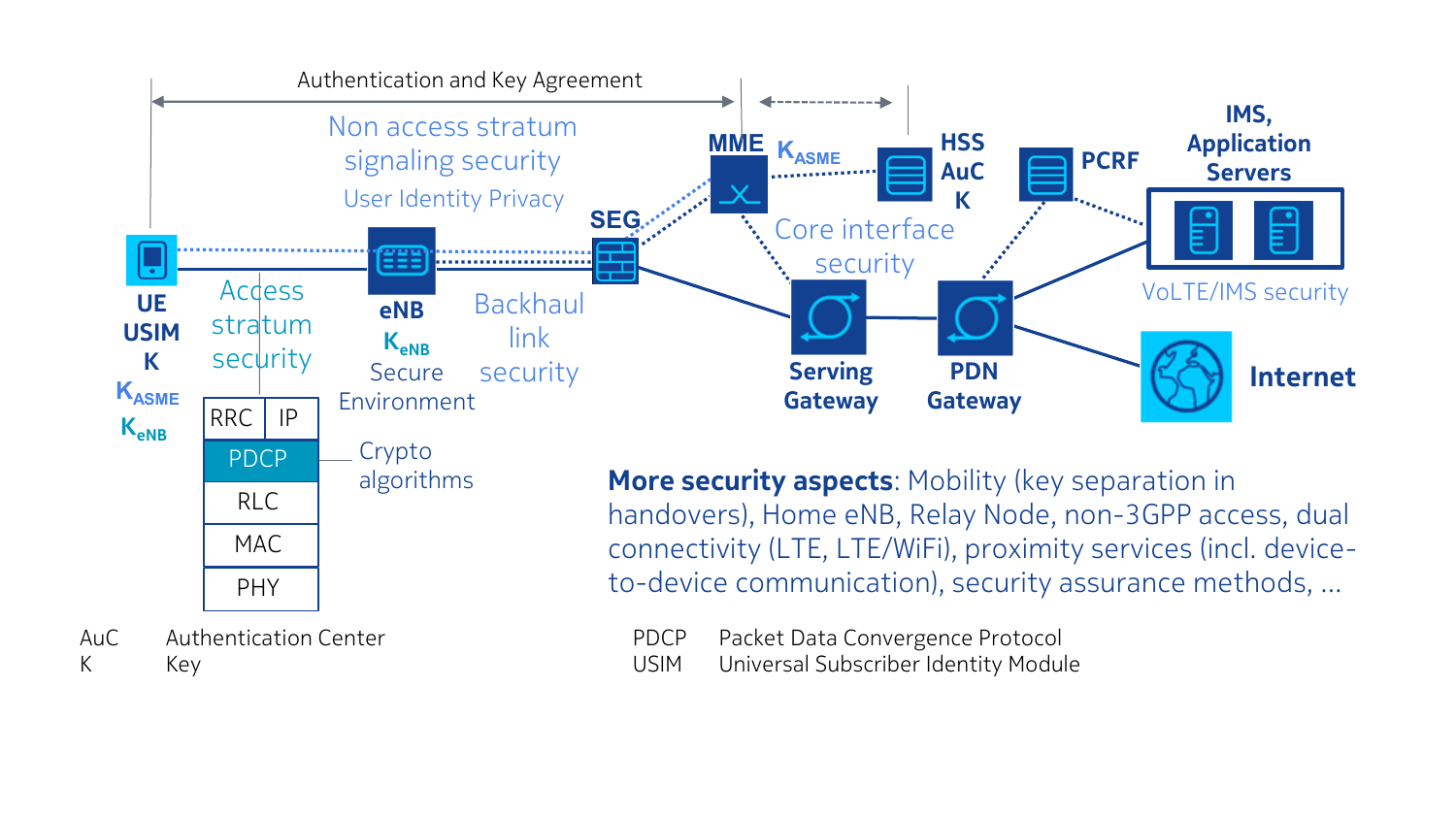}
	{Security Architecture of 3GPP LTE / EPS\cite{5gn2017network}\label{fig:LTE-sec}}
	
	LTE specifies three pairs of \change{cryptographic} algorithms, each pair comprises a confidentiality protection algorithm and an integrity protection algorithm. \change{This redundancy provides fallback options in case a pair} fails during the expected system life-cycle. Also, the backhaul link is protected by Internet key exchange (IKE) / Internet protocol security (IPsec). Generally, the IPsec tunnels are terminated by a dedicated security gateway (SEG), but they may also be terminated at the MME or serving gateway, for the control and user plane, respectively. 
	
	\change{Since the} eNB is considered as a physically exposed entity, with a notable risk of being compromised by an attacker with physical access, \change{it has no access to the NAS signaling.} However, it has access to the user plane traffic that is decrypted and re-encrypted at the eNB between radio interface and backhaul link. \change{Also, 3GPP requires all network elements installed in a ``secure environment'', unauthorized access to eNBs is strictly prohibited. The eNB is also required to protect the keys and other important processes. }
	
	For an overall security concept for an LTE network, additional non-standardized network security measures must be implemented, \change{among them traffic filtering and separation between network-internal security zones, secure operation and maintenance (O\&M), and secure operation of services such like DNS, NTP, IP routing, etc.}
	
	\subsection{Challenges in 5G Multi-Access Edge Computing}\label{subsec:challenges}
	The 3GPP security architecture specified for LTE/EPS cannot fulfill new requirements of emerging services in 5G. 5G networks requires to improve the flexibility of security functions with the MEC services.
	
	The concept of MEC was proposed by the European Telecommunications Standards Institute (ETSI), suggesting to use the base stations instead of centralized cloud servers for offloading computation tasks from mobile users, in order to provide ``\textit{IT and cloud-computing capabilities within the radio access network (RAN) in close proximity to mobile subscribers}''~\cite{etsi2014mobile}.
	Differing from traditional mobile cloud computing systems, MEC systems with hierarchical controlling and management are more near to the end users, less coupled with the central cloud, less dependent on the backhaul network, and therefore with a better support to latency-critical applications~\cite{mao2017survey}. The exploitation of MEC, however, can be constrained by the centralized 3GPP LTE/EPS security architecture.
	
	The current AKA procedure deeply relies on the MME in central cloud, the backhaul latency in network attachment and mobility management can become a bottleneck for the quality of MEC services. When data congestions occur in the core network, this delay will significantly increase, and may violate the service latency requirement, especially for some ultra-reliable and ultra-low-latency emergency communication applications~\cite{metis2015updated}.
	
	Additionally, the emerging concept of ``5G Islands'' \cite{han2018island,kochems2018ammcoa} might be expected 
	 \change{to autonomously provide local devices with central-cloud-independent MEC services.}
	\change{On one hand, 5G Islands can be created in a scheduled manner to serve public events with dense requirement of local MEC traffic, or to maintain MEC services with planned disconnection from the core network. On the other hand, they can be also created upon unexpected demand, so as to provide temporary local MEC services for an improved network resilience against cyber-attacks and extreme disasters (such like fires, explosions, earthquakes etc.) that can cause infrastructural malfunctions .
	This concept is novel, but the development of its elements has already been launched. For example, \cite{simsek2017flexibility} discusses how 5G networks will be able to support highly autonomous and smart decision making capabilities at various layers of the network.
	Network architectures~\cite{yang2016autonomous,jeon2017distributed} were proposed where MMEs autonomously perform UE management in a distributed manner. In this field, distributed \change{and decentralized} edge security methods shall be developed to cope with non-static and unreliable backhaul connections. More specifically, the AKA procedures are expected to be decentralized at the network edge, and to provide stable quality of service of MEC in the 5G.}
	
	\change{Aiming at a decentralized security approach for reliable MEC, there are two key challenges to overcome:
	\begin{itemize}
		\item In the architectural perspective, a novel network architecture is needed to enable a flexible and secure migration of AAA functions between the central cloud and edge clouds in virtualized 5G networks. 
		\item In the operations perspective, it is critical to keep the extra operations cost generated by the decentralization and migration of AAA functions on a reasonable level.
	\end{itemize}
	}
	
	\section{Decentralized Security Approach\\ over Edge Clouds}\label{sec:architecture}
	\subsection{5G Authentication Authorization Accounting}\label{subsec:5gaaa}
	Towards a flexible decentralized AKA mechanism for reliable 5G MEC services, we proposed a novel virtualized AAA approach: the 5G AAA~\cite{stan2017v-aaa}. It combines two independent international standard systems that are the 3GPP and the ETSI, as a single platform to manage and secure the subscribers, the tenants and the network slices under the 5G flexible network environment.
	
	\Figure[htbp!](topskip=0pt, botskip=0pt, midskip=0pt)[width=.8\textwidth]{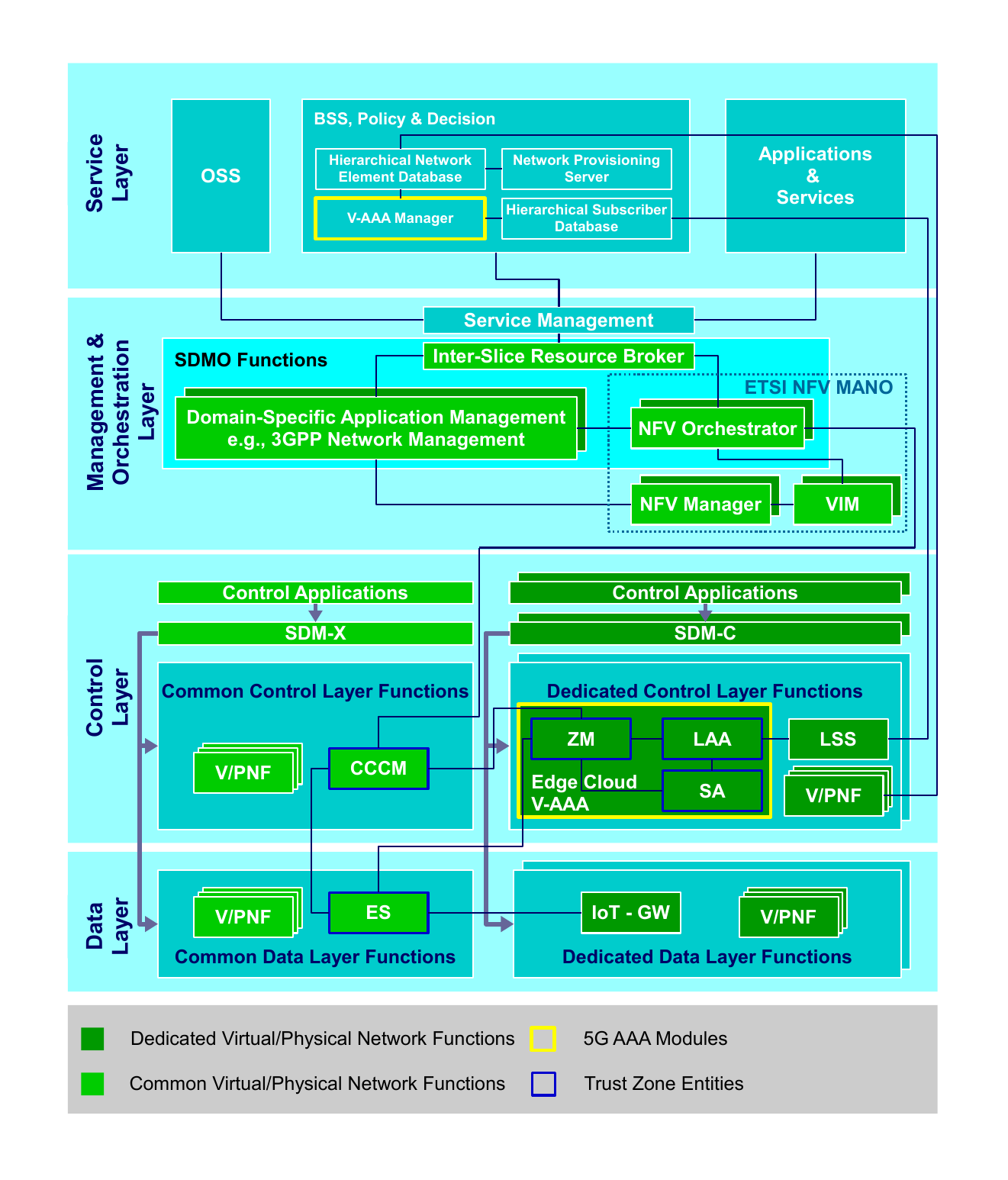}
	{\parbox[t]{.91\textwidth}{The 5G AAA framework enhanced with Trust Zone in the \textit{5G PPP} architecture. The terms OSS, VIM, SDM-X, SDMO and SDM-C stand for operations support system, virtual infrastructure manager, software-defined mobile network (SDM) controller, SDM orchestrator and SDM coordinator, respectively. V/PNF denote virtual/physical network functions.}\label{fig:5gaaa_tz}}
	
	Generally, 5G AAA system is constructed by a \change{V}irtualized-AAA (V-AAA) manager in core network and several V-AAA servers distributed in various ECs, and supported by hierarchical databases approaches for tenant data and tenant's subscribers synchronization. This hierarchical database approach has been divided into two levels of synchronization. Within the tenant network or dedicated resources of tenant network slice would have bi-directional replication across different regions. On the other hand, the tenant's subscriber database would be applied many-to-one synchronization with the MNO core network subscriber database. This a directional synchronization. This logical illustration cab be found in Fig.~\ref{fig:5gaaa_tz}. From the 3GPP view on 5G network architecture \cite{3gpp2017system}, the 5G AAA system is an advanced implementation of the access and mobility management function (AMF), which provides a complete generic AMF set in the central cloud with the V-AAA Manager, while offering specifically customized subsets of AMF for various 5G network slices in different local ECs with the polymorphic V-AAA servers. Similarly, the hierarchical databases and the Local Subscriber Servers (LSS) together construct the 3GPP unified data management (UDM). Taking the 4-layer-view on 5G network architecture proposed by the 5G Infrastructure Public Private Partnership (5G PPP)~\cite{5gppp2016white}, the V-AAA Manager can be seen as part of the business supporting system (BSS) on the service layer, and each EC V-AAA server is a virtual network function (VNF) dedicated to a particular network slice.  
	
	With such an architecture that distributes security functions and databases to all ECs, the 5G AAA approach enables a flexible and decentralized decision and application of security policies in every EC. It is able to:
	\begin{itemize}
		\item Keep the centralized governance of tenants and mobile subscribers in the central cloud;
		\item Allow a degree of freedom of tenant governance their subscribers at the edge cloud;
		\item Use the AKA generated data to track and authenticate the mobile subscribers from the edge cloud and central cloud;
		\item Maintain the mobile subscribers and tenants with the point of attachment in the central cloud as well as in the edge cloud;
		\item \change{Retain} subscriber services even when the backhaul connection is limited or unavailable;
		\item Provide a trust platform collecting the trust value from the mobile subscribers to the NFV-based tenants, and from the physical network entities to the virtual network entities.
	\end{itemize}

	\subsection{Trust Zone}\label{subsec:tz}
	The proposed 5G AAA \cite{stan2017v-aaa} approach enables the AKA procedure in the local EC to improve the MEC services. Nevertheless, its deployment still calls for novel solutions of adaptive AMF invocation. Due to the incomplete AMF set and probably limited budget on security measures, the local V-AAA servers are usually less secure than the central V-AAA Managers. Therefore, it is generally preferred to invoke the AMF from the central cloud in common cases, and only to rely on the local version when the backhaul connection cannot satisfy the latency requirement of MEC services due to -- intentional or unintentional -- restrictions, congestions and collapses. To achieve this, the 5G AAA system must be able to adaptively switch between the centralized and local security modes with respect to the current backhaul connection status. The switching process must be carefully designed to mitigate data leaks.
	
	Motivated by this challenge, we designed an EC security architecture, the Trust Zone (TZ), in order to enhance and extend the 5G AAA approach to a cognitive solution~\cite{han2017security}. The basic principle of TZ is to let the local EC cognitively measure the quality of its backhaul connection. When the backhaul is healthy, the AMF is invoked at the V-AAA Manager in central cloud. When the backhaul connection is seriously limited or unavailable, the TZ relies on the V-AAA server that executes local AMF to maintain MEC services.
	
	In a structural view, as illustrated in Fig.~\ref{fig:5gaaa_tz}, a TZ system consists of five entities: 
	\begin{enumerate}
		\item \textbf{Central Cloud Connection Monitoring (CCCM)} --
		This entity periodically visits the NFV orchestrator in central cloud to evaluate the backhaul status. Additionally, when the backhaul connection is limited or unavailable, CCCM tries to diagnose the malfunction. This \change{diagnosis, including information such the likely location and reason of malfunction,} can be used by the software-defined network (SDN) management to assist backhaul recovery with backup resources, e.g. network redundancy and alternative routes such as satellite links. A dynamic network resource allocation in the SDN management can also be supported by the backhaul status information. When the backhaul connection becomes poor, specific network functions such as mobility management and fault management can obtain network resources with higher priorities.
		\item \textbf{Zone Management (ZM)} --
		This is the core entity of TZ, which implements \change{most of AMFs} and is connected with all other TZ entities. It triggers and coordinates the state transitions in TZ, when it receives reports about updated backhaul status. Integrated with interfaces to the central cloud and to the UEs, it also collaborates with the Local Access Assistant (LAA) to accomplish the AS security procedures in the local EC. When the backhaul connection is healthy, ZM cooperates with the V-AAA Manager in central cloud to execute centralized AKA. When the backhaul is limited or unavailable, local authentication is needed, ZM will then collaborate with the LAA to execute local AKA instead. 
		\item \textbf{Local Access Assistant (LAA)} --
		This entity makes use of the subscriber information that is essential for authentication and stored in the LSS, in order to locally execute security measures such as AKA procedures. It can be considered as a lite AMF module migrated from the central cloud to the local EC. LAA is deactivated by the ZM when the backhaul is reliably healthy, and activated otherwise. The LAA is isolated from the ZM mainly due to security concerns: the subscriber database should be decoupled from the ZM that is the central controlling unit of TZ and the priori target of potential cyber-attacks. To mitigate the risk that the LAA is attacked and manipulated to send fake information to the central cloud, an asymmetric trust relation between the central V-AAA manager and the local V-AAA server is needed, as will be introduced in Sec. \ref{subsec:asymmetric_trust}. 
		\item \textbf{Security Auditing (SA)} --
		Mobile network operators (MNOs) usually keep logs of security critical operations for each subscriber. These logs can be audited to investigate all potential risks of illegal access and cyber-attacks. Both the log database and the auditing center are usually located in the central cloud, so when the backhaul connection is unavailable, they cannot keep tracking the AAA operations executed by local V-AAA servers. To close this gap and provide a continuous audit, the SA module is designed. It is activated under the local security mode to record all local AAA operations. After the backhaul recovery, these records can be either actively pushed to the central auditing center, or passively pulled upon request.
		\item \textbf{Emergency Services (ES)} --
		we have named some extreme disasters that can lead to backhaul outages in Sec. \ref{subsec:challenges}, such as fires, explosions and earthquakes. These events can be harmful not only to the network infrastructure, but also to human safety. A set of emergency services can be defined, which help users avoid personal injury and property damage under such disasters, even when their devices cannot be authenticated by the network. These services are implemented in the ES entity, including but not limited to public disaster alarm, evacuation guidance, positioning service, emergency call and short message service (SMS). Additionally, the ES is connected to the Internet-of-things gateway (IoT-GW) to collect context information about public disasters from external sensor networks. Such information can be visited by the network function virtualization management and orchestration (NFV MANO) through the CCCM to assist the network diagnose and backhaul recovery processes.
	\end{enumerate}
	Among those entities listed above, the first three entities are dedicated to a specific tenant or network slice, and within the EC V-AAA server. However, CCCM and ES are defined as common functions and non-V-AAA server functions exist on both control plane and user plane\footnote{Further details about the interfaces between TZ entities and a behavior model of TZ are available in \cite{han2017security}.}
	
	\subsection{Asymmetric Trust to Safely Transfer Access Management}\label{subsec:asymmetric_trust}
	It should be noted that local V-AAA servers are less secure than the central V-AAA Manager\change{. Cyber-attackers} may initiate attacks to disconnect the central cloud and the EC, and hack the local TZ that is easier to target than the central cloud. Then they may eventually attempt to obtain access to the central cloud during the backhaul recovery, when the TZ hands the access management back over to the central cloud. To mitigate this risk, we designed an asymmetric approach of transferring the access management between the central cloud and the EC, which can be briefly summarized as follows: When a degradation of backhaul connection occurs and the TZ switches to the local security mode, all devices already authenticated by the central V-AAA Manager before the mode switching will be considered as trusted by the local V-AAA server, and retain a seamless access to the local MEC services. In contrast, devices that newly arrive after the switching must go through a local authentication and authorization process to access the MEC services. Only the users that are successfully authenticated and authorized are considered as temporarily trusted to access the local MEC services, while for the rest devices only the emergency services will be available. When the backhaul connection recovers, all temporarily trusted devices in the local EC have to reconnect to the network in a preplanned order, so that they are authenticated and authorized by the central V-AAA Manager again.
	
	\section{Context-Awareness in Trust Zone}
	\subsection{Context-Aware Synchronization of Subscriber Profiles}\label{subsec:sync}
	The proposed local V-AAA server -- or more specifically, the LAA -- relies on the subscriber authentication information stored in the LSS to accomplish the local AKA procedure in the EC. However, due to security concerns, parts of such information, such like the NAS key, cannot be locally generated in ECs, but only available in the central cloud. Thus, the subscriber authentication information must be synchronized from the central cloud to the LSS before the EC switches to the local security mode, and periodically updated. Furthermore, taking the device mobility into account, users may enter an EC from \change{outside} while it is \change{suffering from a high backhaul latency -- or in the worst case, }disconnected from the central cloud. Therefore, a LSS should not only synchronize the authentication information of devices in the local EC, but also that of the devices in neighbor ECs nearby, which are able to arrive in the local EC during the next synchronization interval. 
	
	Self-evidently, with respect to the mobility, various devices can move over different distances within the same synchronization interval. Thanks to the modern technologies of transportations and communications, people are nowadays able to travel over three hundred kilometers per hour, while continuously exploiting the cellular network services. Considering the limited area of a typical EC, e.g. between $5\times5~\text{km}^2$ and $15\times15~\text{km}^2$ \cite{checko2015cloud}, the authentication information of every such high-mobility subscriber must be synchronized to 400 to 4000 different LSSs per hour. Accounting the large amount of user devices in the entire mobile network, this synchronization process will generate a considerably significant data traffic over the backhaul network, and hence leads to extra operating cost and raises the risk of data congestion. In other words, there is a conflict between the interests of subscribers (MEC service reliability) and of MNOs (operating cost). In this concern, we designed an advanced synchronizing mechanism with awareness \change{of} the context information of both UEs and ECs, which can be explained as follows.
	
	The local security mode is activated only when the central security services fail to respond in time, the gain of subscriber authentication information synchronization is therefore not significant in the ECs with negligible probability of Central Security Service Outage (CSSO). \change{Note that we refer to the term CSSO as the cases where the security network functions in central cloud fail to respond within the desired latency, instead of a complete unavailability of service.} Generally, to estimate this outage probability for a certain device in different ECs, the following factors should be taken into account:
	\begin{enumerate}
		\item \textbf{User arrival}: depending on the device-to-edge-cloud spatial distance and the user mobility, every EC has its individual expectation about the user arrival probability in the next synchronization interval. The ECs located far from the device have less chance to serve the device than those located nearby. For still and low mobility users, the user arrival estimation can be simply accomplished though the handover prediction procedure of legacy networks. However, high-mobility users such like vehicles on highway and railway passengers can travel over multiple ECs during one backhaul outage, so that an additional position prediction in macro scale can be \change{necessary}.
		
		\item \textbf{User stay time}: upon arrival, depending on the user mobility, the coverage area of the EC and the traffic model in that area, every EC has its individual expectation about the duration of user's stay in it. Some ``important'' ECs covering larger or more traffic-congested areas (e.g. the city center of Paris) are expected to serve a user for longer time, when compared to those ``unimportant'' ones (e.g. a small village aside a high-speed railway). The risk of CSSO during the user's stay in these ECs, therefore, is also higher.
		
		\item \textbf{Backhaul reliability}: depending on the current status of backhaul connection, every EC has its individual expectation about the chance that a CSSO occurs during a certain period in the future. Generally, the ECs with worse current backhaul connection quality are more likely to suffer from CSSO.
	\end{enumerate}
	
	Thus, from real-time information of devices (mobility and position) and ECs (coverage area, traffic model and backhaul throughput), we are able to extract high-level context information such as user motion pattern and backhaul outage probability, as illustrated in Fig.~\ref{fig:synchronization}. These context information can be used to estimate \change{the CSSO risk in different ECs for every certain device.}
	
	\change{More specifically, given a certain EC $C$,  let $T$ denote the synchronization interval, the duration of $u$ suffering from CSSO in $C$ in the next period of $T$ is
	\begin{equation}
		\text{E}\{t_{\text{o},u,C}\}=p_{\text{o},C}\int_{0}^{T}f_{\text{arr},u}(t)\int_{0}^{T-t}f_{\text{stay},u}(\tau)\tau\text{d}\tau\text{d}t
	\end{equation}
	where $p_{\text{o},C}$ is the predicted probability of CSSO in $C$ over the next period $T$, $f_{\text{arr},u,C}(t)$ denotes the probability density that $u$ arrives in the coverage of $C$ after $t$, and $f_{\text{stay},u,C}(\tau)$ is the probability density function of $u$'s stay time $\tau$ in the coverage of $C$ upon arrival.	The decision of synchronizing the subscriber authentication information of $u$ to $C$ can be then made following:
	\begin{equation}\label{equ:sync_decision}
		\begin{cases}
			\text{To synchronize} & \text{E}\{t_{\text{o},u,C}\}/T\geq \mu;\\
			\text{Not to synchronize} &\text{E}\{t_{\text{o},u,C}\}/T< \mu,
		\end{cases}
	\end{equation}
	where $\mu$ is a pre-defined threshold,	so that less redundant backhaul traffic is generated. By setting $\mu$ to different levels, the MNO is able to balance its interests in operating cost and MEC service reliability.}
	
	\change{
	Moreover, inspired by \cite{han2018island}, if the MNO is able to evaluate the loss of $u$ in suffering from CSSO per unit time, call it $l_u$, and the operations cost $c_{\text{s},C}$ to synchronize the subscriber authentication information of a UE to $C$, there is a cost-optimal strategy that minimizes the sum of potential CSSO loss and operations cost:
	\begin{equation}
		\begin{cases}
			\text{To synchronize} & \text{E}\{t_{\text{o},u,C}\}\cdot l\ge c_{\text{s},C};\\
			\text{Not to synchronize} &\text{E}\{t_{\text{o},u,C}\}\cdot l< c_{\text{s},C}.
		\end{cases}
	\end{equation}
	Therefore, we can individually solve the optimal threshold for every $C$:
	\begin{equation}\label{eq:opt_threshold}
		\mu_{\text{opt},C}=\frac{c_{\text{s},C}}{lT}.
	\end{equation}
    }

	\Figure[htbp!](topskip=0pt, botskip=0pt, midskip=0pt)[width=.48\textwidth]{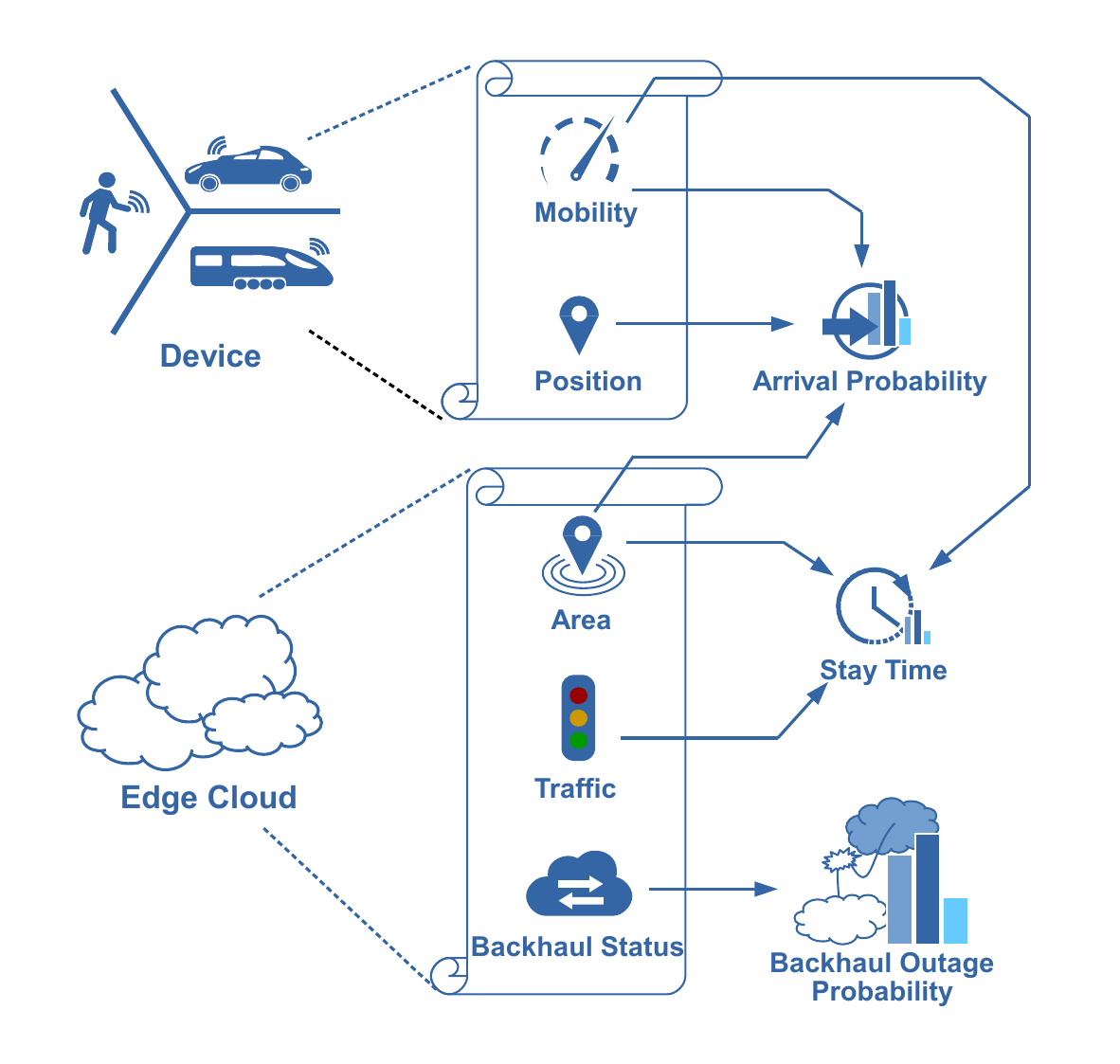}
	{Trust Zone extracts the context information of UEs \& EC to estimate the central security service outage probability.\label{fig:synchronization}}	

	\subsection{Modeling the Backhaul Connection Status}\label{sec:model}
	Although mature models and techniques are available to estimate user \change{mobility} in mobile networks~\cite{ge2016user}, which can be applied in the CSSO estimation procedure discussed above, a simple and efficient model for the backhaul status is still called for.
	
	Generally, a backhaul connection between the central cloud and the edge cloud can be evaluated with its average delay and packet error rate, which are easy to measure. With these metrics, we are able to estimate the probability that the central security server responses in time to support a certain edge computing service, when given its specific latency requirement. We define this probability as the Central Security Service Reliability (CSSR), which exhibits stochastic behavior in long term and can be modeled as a Markov chain \change{according to \cite{pukite1998markov}}, as shown in Fig.~\ref{fig:cssr_as_markov}.
	\Figure[htbp!](topskip=0pt, botskip=0pt, midskip=0pt)[width=.48\textwidth]{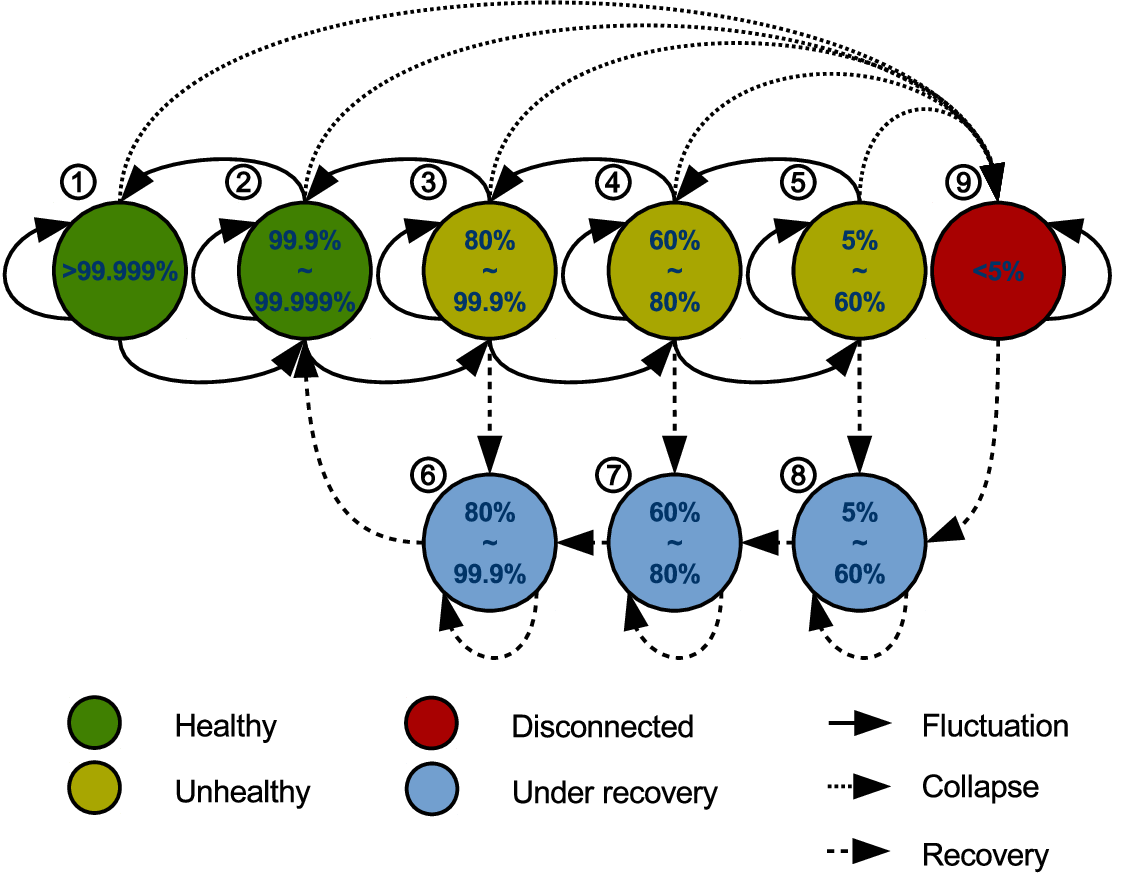}
	{Central security service reliability as a Markov chain, circled numbers are state indexes.\label{fig:cssr_as_markov}}	
	
	In our model, the backhaul network at any time instant is considered to be in one among several predefined states. Every state describes a specific combination of CSSR range and network state. According to the CSSR level, we defined three basic network states: \textit{healthy}, \textit{unhealthy} and \textit{disconnected}. Additionally, the network operator may recognize network disasters and take necessary measures to repair the backhaul connection, for which we set an extra state \textit{under recovery}. 
	
	Usually, the CSSR does not vary fiercely but smoothly, so only fluctuations between neighbor CSSR levels are considered, as we illustrate with solid arrows in Fig.~\ref{fig:cssr_as_markov}. Nevertheless, as mentioned in Sec.~\ref{subsec:challenges}, some disasters and cyber-attacks, although rarely happen, can cause sudden disconnection of \change{backhaul}. In this case, the backhaul connection will fail to autonomously recover by itself, unless manual repair is taken. Therefore, in the model we enabled unidirectional transitions from the \textit{healthy} and \textit{unhealthy} states to the \textit{disconnected} state, which are illustrated with dotted arrows in Fig.~\ref{fig:cssr_as_markov}. Furthermore, the network operator will take autonomous or manual measures to recover the backhaul network only when it is not healthy, and the recovery process will continuously improve the CSSR until it becomes healthy again, as denoted with the dashed arrows in Fig.~\ref{fig:cssr_as_markov}. By training the model with historical log data of the network operator, we are able to estimate the transition probabilities and predict the future CSSR. \change{A demonstrative example} output of our Markov chain is depicted in Fig.~\ref{fig:cssr_as_markov_example}, \change{which is generated from an artificial state transition probability matrix $\mathbf{T}$ of
		\begin{equation*}
		{\fontsize{6}{0}
		\begin{bmatrix}
		0.8&0.1999&0&0&0&0&0&0&0.0001\\
		0.5&0.49&0.0099&0&0&0&0&0&0.0001\\
		0&0.25&0.5&0.15&0&0.0999&0&0&0.0001\\
		0&0&0.25&0.5&0.15&0&0.0999&0&0.0001\\
		0&0&0&0.2&0.6&0&0&0.1&0.1\\
		0&0.5&0&0&0&0.5&0&0&0\\
		0&0&0&0&0&0.5&0.5&0&0\\
		0&0&0&0&0&0&0.5&0.5&0\\
		0&0&0&0&0&0&0&0.3&0.7\\
		\end{bmatrix}}.
		\end{equation*}
		Here, $T_{i,j}$ denotes the probability of transition from the state $i$ to $j$, for all $\{i,j\}\in\{1,2,\dots,9\}^2$.}
		\Figure[htbp!](topskip=0pt, botskip=0pt, midskip=0pt)[width=.48\textwidth]{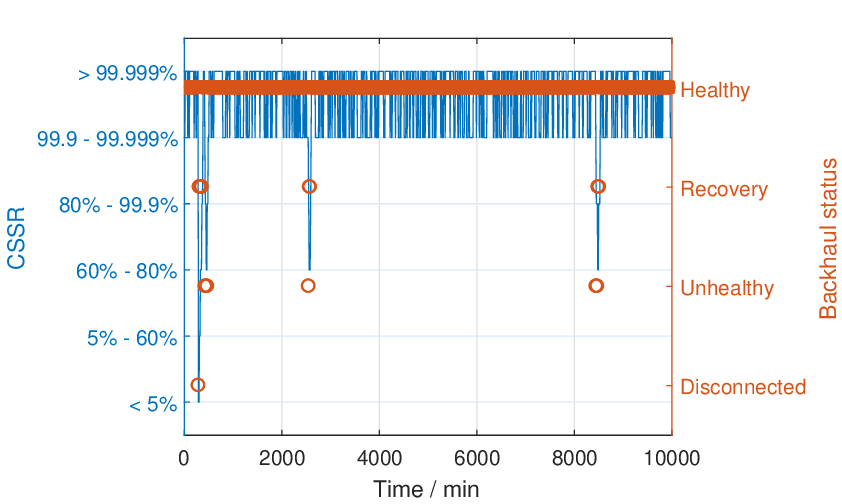}
		{\change{An example simulation of CSSR generated by our Markov chain (the blue curve), and the corresponding network state (the red circles).}\label{fig:cssr_as_markov_example}}	
	
	\section{Numerical Simulation}\label{sec:simulation}
	We conduct numerical simulation to validate the effectiveness of the TZ approach, as to describe below.
	\subsection{Environment Setup}
	An $8\times 8~\text{km}^2$ region with UE density of $6250/\text{km}^2$ is considered. An EC covers a circular urban area with radius of $2~\text{km}$ that locates in the center of the region, which is surrounded by four suburban and rural areas, as illustrated in Fig. \ref{fig:sim_map}.  
	\Figure[htbp!](topskip=0pt, botskip=0pt, midskip=0pt)[width=.2\textwidth]{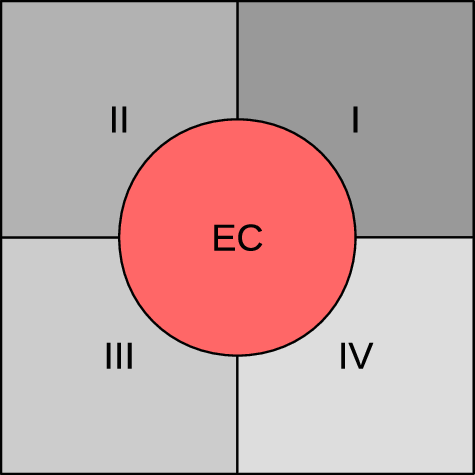}
	{Geographical map of the simulated region, EC is the coverage of the local EC, while I -- IV are neighbor suburban and rural areas.\label{fig:sim_map}}	
	
	The simulation initializes with number of UEs that are uniformly distributed across the region. \change{
	Also, every UE is assigned to a mobility class that is randomly selected from \emph{HIGH}, \emph{MEDIUM}, \emph{LOW} and \emph{STILL} with equal probabilities.
	To generate the mobility of individual users, a variety of models have been proposed that can be classified into four categories, namely the random walk model, the random waypoint	model, the fluid flow model and Gauss-Markov model, respectively\cite{ge2016user}. Considering the dependency of user mobility on the terrain and traffic environment, here we have extended the random walk model with $600~\text{s}$ steps, where a Gaussian distributed} ``basic speed'' is generated for every moving UE according to its mobility class. This basic speed is then linearly scaled according to both the mobility class and the area where the UE is located, and combined with an uniformly distributed random direction, in order to obtain the UE's movement. Details are listed in Tabs. \ref{tab:basic_speed} and \ref{tab:speed_correction}. Once a UE leaves the region, a new arriving UE will be created to keep a constant overall UE density. We let the UEs move for $40~\text{h}$ so that the UE densities in all five areas converge to consistent levels.
	\begin{table}[!h]
		\centering
		\caption{Basic speed of different UE mobility classes (in m/s)}
		\label{tab:basic_speed}
		\begin{tabular}{c*{4}{|c}}
			\toprule[2px]
			\scriptsize Mobility Class & \scriptsize\emph{STILL} & \scriptsize\emph{LOW} & \scriptsize\emph{MEDIUM} & \scriptsize\emph{HIGH}\\\hline
			\scriptsize Basic Speed & \scriptsize 0 & \scriptsize $\sim\mathcal{N}(1.5,0.5)$ &  \scriptsize$\sim\mathcal{N}(10,2)$ &  \scriptsize$\sim\mathcal{N}(40,5)$\\
			\bottomrule[2px]
		\end{tabular}
	\end{table}
	\begin{table}[!h]
		\centering
		\caption{Speed scaling factors for different mobility classes and areas, still UEs omitted as they do not move}
		\label{tab:speed_correction}
		\begin{tabular}{c*{5}{|c}}
			\toprule[2px]
			\diagbox{Mobility Class}{Area} 	& EC & I & II & III & IV\\\hline
			\emph{LOW} 								   & 1 	  & 1 & 1 & 1 & 1\\\hline
			\emph{MEDIUM} 							& 0.7 & 1 & 0.9 & 0.8 & 0.9\\\hline
			\emph{HIGH} 								& 0.2 & 1 & 0.9 & 0.85 & 0.8\\
			\bottomrule[2px]
		\end{tabular}
	\end{table}
	
	Meanwhile, we take the Markov chain described in Sec. \ref{sec:model} to generate the CSSR of unstable central security services in the EC as a random process, which is updated by every simulation step. In every simulation step, depending on the current CSSR, there is a specific chance that a CSSO occurs in the EC.
	
	\subsection{Testing the Trust Zone}
	We consider a TZ implemented in the EC serving an urban area. It is assumed to have complete knowledge about the Markov chain parameters of the EC's CSSR. After the convergence of UE densities in different areas, the TZ tracks the positions of UEs for $24~\text{h}$ to learn about the statistical motion patterns, in order to predict the arrival probability and stay time of every UE in the EC. After the training phase, the TZ is activated to estimate the \change{CSSO risk criterion $\text{E}\{t_{\text{o},u,C}\}/T$ in the next $T=30~\text{min}$ for every UE $u$ in the region}. If and only if the risk exceeds a pre-defined threshold, the UE's subscriber profile will be synchronized from the central cloud to enable TZ operation upon CSSO, and a backhaul traffic will be thereby generated. This operation repeats every $30~\text{min}$. By every simulation step, if a CSSO occurs in the EC, every UE that has not had its subscriber profile synchronized will suffer from the CSSO once, and therewith generate a CSSO report. \change{To demonstrate the potential of cost optimization indicated by \eqref{eq:opt_threshold}, we assume a normalized loss rate $l=1/\text{hour}$ and $c_{\text{s},C}=5\times10^{-5}$ so that $\mu_{\text{opt},C}=1\times10^{-4}$.} This testing phase continues one month, and the performance of TZ under different specifications of \change{decision threshold} are illustrated in Fig. \ref{fig:sim_results}.
	
	It can be observed that the deployment of TZ can effectively reduce the loss of user experience caused by CSSOs. Especially, by configuring the decision threshold to different levels, we can achieve a flexible balance between the service reliability and the operations cost. \change{Especially, the sum of CSSO-caused loss and operations cost is minimized at the analytical optimum of decision threshold.}
	\Figure[htbp!](topskip=0pt, botskip=0pt, midskip=0pt)[width=.47\textwidth]{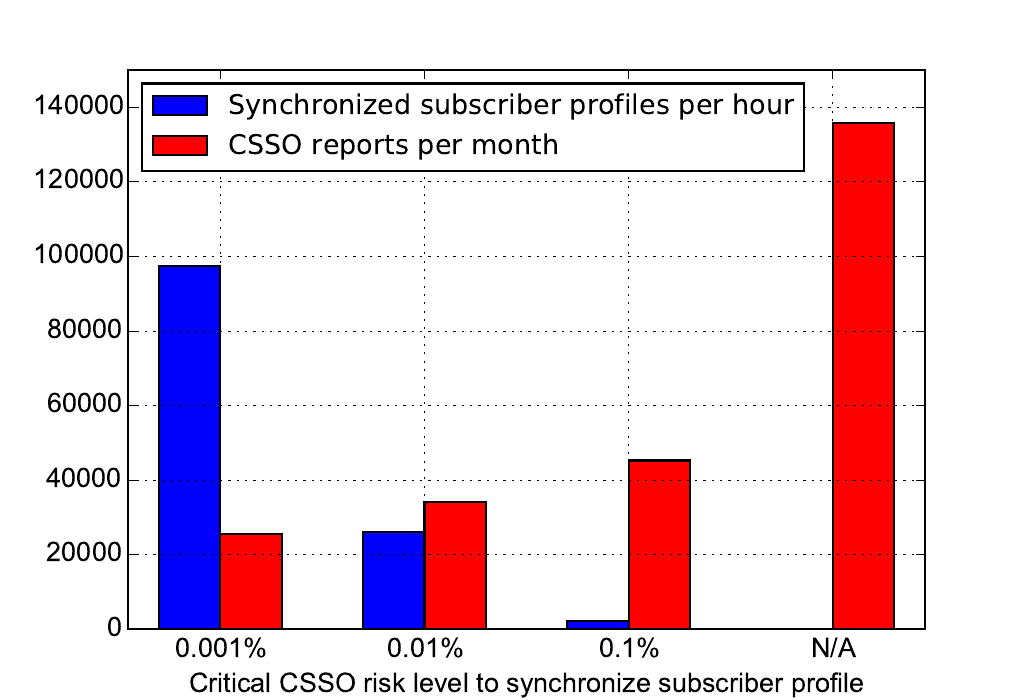}
	{The performance of TZ under different specifications of \change{decision threshold. Note that N/A stands for no TZ deployment, which is the baseline of LTE systems.}\label{fig:sim_results}}
	
	\section{Conclusion and Further Challenges}\label{sec:conclusion}
	In this paper, we have indicated the future 5G telecommunication networks with dense application of MEC services, and the current LTE/EPS security architecture in 3GPP Release 14 drawbacks with its centralized AKA mechanism. We proposed a solution integrated with our 5G AAA approach which enables the execution of AKA \change{functionality} at the ECs, in order to support better deployment of 5G MEC services. An innovative EC security architecture, the Trust Zone, has been introduced to enhance the 5G AAA with a cognitive access management with respect to the backhaul connection quality. Furthermore, we designed a context-aware mechanism of synchronizing subscriber authentication information to local subscriber databases, which helps in reducing the backhaul network traffic generated by the Trust Zone. Numerical simulation showed that our proposed method can effectively improve the reliability of EC services in the context of unstable security network functions in central cloud, while being capable to keep a flexible balance between the MEC reliability and the operations cost.
	
	We will continue to investigate better methods to improve the authentication process in 5G. \change{It shall be noted that some assumptions and simplifications, such as the uniform distribution of \change{UE} and the extended random walk model, can be inaccurate in practical use cases. Advanced models such as the work reported in \cite{ye2018} will be applied to simulate more realistic user mobility.} In terms of increasing the service reliability, we will apply machine learning and deep learning techniques \change{to build precise models from field measurements of CSSR, in order} to improve the accuracy and efficiency of centralized security. Furthermore, the proposed decentralized security mechanisms need to be evaluated with respect to their applicability of local authentication and authorization in hybrid private-public mobile networks as expected for 5G, e.g. 5G enterprise networks inter-working with 5G public land mobile networks.

   \vfill
   \begin{IEEEbiography}[{\includegraphics[width=1in,height=1.25in,clip,keepaspectratio]{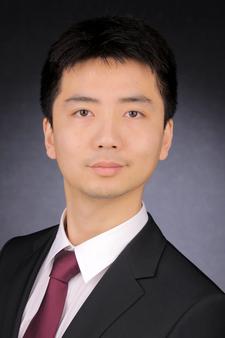}}]{Bin Han}
   	received his B.E. degree in 2009 from Shanghai Jiao Tong University, China, and his M.Sc. degree in 2012 from Technische Universit\"at Darmstadt, Germany. In 2016 he was granted the Ph.D. (Dr.-Ing.) degree from Kalsruhe Institute of Technology, Germany. He joined the Institute of Wireless Communication, Technische Universit\"at Kaiserslautern, Germany, since July 2016, and is currently working as senior lecturer. His research interests are in the broad area of communication systems and signal processing\change{, with a recent special focus on MEC and 5G network slicing}.
   \end{IEEEbiography}
   
   \vfill
   
   \begin{IEEEbiography}
   	[{\includegraphics[width=1in,height=1.25in,clip,keepaspectratio]{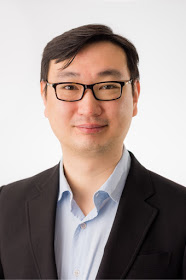}}]{Stan Wong} received BEng and M.Sc. from King's College London (KCL), and Ph.D. from University of London in 2009. In 2014, he joined an IT consultancy firm in Hong Kong as a system consultant and responsible for metro Ethernet 2.0 network infrastructure design. In 2015, he rejoined KCL and worked on the EU-funded project covering the 5G Security. In 2016, he won the collaborative spectrum sharing prize from European Commission. He has deep knowledge in 5G security, machine learning data analytic, and different standard IoT air interfaces and application. \change{In 2018, he joined GSM Association (GSMA) and lead the 5G Security in GSMA. }
   \end{IEEEbiography}
   
   
   \begin{IEEEbiography}
   	[{\includegraphics[width=1in,height=1.25in,clip,keepaspectratio]{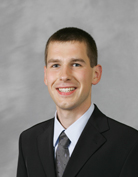}}]
   	{Christian Mannweiler} received his M.Sc. (Dipl.-Wirtsch.-Ing.) and Ph.D. (Dr.-Ing.) degrees from the Technische Universit\"at Kaiserslautern (Germany) in 2008 and 2014, respectively. Since 2015, he is a member of the Cognitive Network Management research group at Nokia Bell Labs, where he has been working in the area of network management automation and SON for 5G systems. He is (co-)author of numerous articles and papers on wireless communication technologies and architectures for future mobile networks. Christian has worked in several nationally and EU-funded projects covering the development of cellular and industrial communication systems.
   \end{IEEEbiography}
   
   \vfill
   
   \begin{IEEEbiography}[{\includegraphics[width=1in,height=1.25in,clip,keepaspectratio]{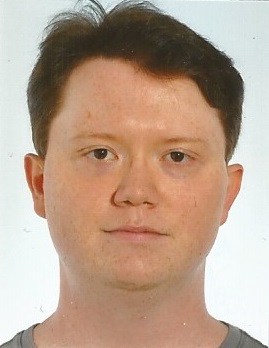}}]{Marcos Rates Crippa} received his B.Sc. degree in Computer Science in 2010 from the Federal University of Rio Grande do Sul, Brazil, and his M.Sc. degree in 2013 from the Technische Universit\"at Kaiserslautern, Germany. Since 2013, he has a position at the Institute of Wireless Communication (WiCon), Technische Universit\"at Kaiserslautern, Germany, as research associate. His main interests are in Network Function Virtualization and cellular networks architecture. He has worked in German and EU-funded project in the areas of spectrum sharing, SDN, NFV and mobile network architectures.
   \end{IEEEbiography}
   
	\vfill
   
   \begin{IEEEbiography}[{\includegraphics[width=1in,height=1.25in,clip,keepaspectratio]{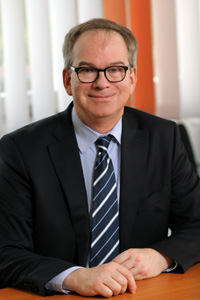}}]{Hans D. Schotten} received the Diplom and Ph.D. degrees in Electrical Engineering from the Aachen University of Technology RWTH, Germany in 1990 and 1997, respectively. Since August 2007, he has been full professor and head of the Institute of Wireless Communication at the Technische Universit\"at Kaiserslautern. Since 2012, he has additionally been Scientific Director at the German Research Center for Artificial Intelligence heading the ``Intelligent Networks'' department.
   \end{IEEEbiography}

	
	\EOD
	

\begin{thebibliography}{1}
		\bibitem{barbarossa2014communicating}
		Sergio Barbarossa, Stefania Sardellitti, and Paolo Di Lorenzo. ``Communicating While Computing: Distributed Mobile Cloud Computing over 5G Heterogeneous Networks'', \emph{IEEE Signal Processing Magazine} 31.6 (2014): 45-55. 
		
		\bibitem{sabella2016mobile}
		Dario Sabella, et al. ``Mobile-edge computing architecture: The Role of MEC in the Internet of Things'', \emph{IEEE Consumer Electronics Magazine} 5.4 (2016): 84-91.
		
		\bibitem{corcoran2016mobile}
		Peter Corcoran, and Soumya Kanti Datta. ``Mobile-Edge Computing and the Internet of Things for Consumers: Extending Cloud Computing and Services to the Edge of the network'', \emph{IEEE Consumer Electronics Magazine} 5.4 (2016): 73-74.	
		
		\bibitem{mao2017survey}
		Yuyi Mao, Changsheng You, Jun Zhang, Kaibin Huang and Khaled B. Lataief, ``A Survey on Mobile Edge Computing: The Communication Perspective'', \emph{IEEE Communications Surveys \& Tutorials}. IEEE, 2017.
		
		\bibitem{abbas2018mobile}
		Nasir Abbas, Yan Zhang, Amir Taherkordi, et al. ``Mobile Edge Computing: A Survey'', \emph{IEEE Internet of Things Journal} 5.1 (2018): 450-465.
		
		\bibitem{lal2017nfv}
		Shankar Lal, Tarik Taleb, and Ashutosh Dutta. ``NFV: Security Threats and Best Practices'', \emph{IEEE Communications Magazine}, Vol. 55, Issue 8, pp. 211--217. IEEE, 2017.
		
		\bibitem{li2018security}
		Hongwei Li, Rongxing Lu, Jelena Misic, et al., ``Security and Privacy of Connected Vehicular Cloud Computing'', \emph{IEEE Network}, Vol. 32, Issue 3, pp. 4-6. IEEE, June 2018
		
		\bibitem{klein2010access}
		A. Klein, C. Mannweiler, J. Schneider and H. D. Schotten, ``Access Schemes for Mobile Cloud Computing'', in \emph{2010 Eleventh International Conference on Mobile Data Management}, Kansas City, MO, USA, 2010, pp. 387-392.
		
		\bibitem{3gpp-33401}
		3GPP Technical Specification Group Services and System Aspects, \emph{3GPP TS 33.401 v15.0.0: 3GPP System Architecture Evolution (SAE); Security Architecture (Release 15)}, June 2017.
		
		\bibitem{3gpp-33402}
		3GPP Technical Specification Group Services and System Aspects, \emph{3GPP TS 33.402 v14.2.0:  3GPP System Architecture Evolution (SAE); Security aspects of non-3GPP accesses (Release 14)}, June 2017 .
		
		\bibitem{5gn2017network}
		Ignacio Labrador Pavon, Jorge Rivas S\'anchez, Alessandro Colazzo, et al., ``5G NORMA Network Architecture -- Intermediate Report'', Technical Report, January 2017.
		
		\bibitem{etsi2014mobile}
		ETSI, \emph{Mobile-Edge Computing -- Introductory Technical White Paper}, White Paper, September 2014.
		
		
		\bibitem{metis2015updated}
		Mobile and Wireless Communications Enablers for the Twenty-Twenty Information Society (METIS), ``Deliverable 1.5: Updated Scenarios, Requirements and KPIs for 5G Mobile and Wireless System with Recommendations for Future Investigations'', \emph{ Deliverable}, ICT-317669-METIS, 2015.
		
		\bibitem{han2018island}
		Bin Han, Marcos Rates Crippa and Hans D. Schotten, ``5G Island for Network Resilience and Autonomous Failsafe Operations'', \emph{2018 European Conference on Networks and Communications (EuCNC)}, Ljubljana, Slovenia, June 2018.
		
		\bibitem{kochems2018ammcoa}
		Jacob Kochems and Hans D. Schotten, ``AMMCOA - Nomadic 5G Private Networks'', \emph{23. VDE/ITG Fachtagung Mobilkommunikation)}, Osnabrück, May 2018
		
		\bibitem{simsek2017flexibility}
		Meryem Simsek, Dan Zhang, David {\"O}hmann, Maximilian Matth{\'e}, and Gerhard Fettweis, ``On the Flexibility and Autonomy of 5G Wireless Networks'', \emph{IEEE Access}. IEEE, 2017.
		
		\bibitem{yang2016autonomous}
		Hua Yang, Naoki Wakamiya, Masayuki Murata, Takanori Iwai, and Satoru Yamano, ``An Autonomous and Distributed Mobility Management Scheme in Mobile Core Networks'', \emph{Proceedings of the 9th EAI International Conference on Bio-inspired Information and Communications Technologies}. IEEE, 2016.
		
		\bibitem{jeon2017distributed}
		Seil Jeon, et al. ``Distributed Mobility Management for the Future Mobile Networks: A Comprehensive Analysis of Key Design Options'', \emph{IEEE Access} 5 (2017): 11423-11436.
		
		\bibitem{stan2017v-aaa}
		Stan Wong, Nishanth Sastry, Oliver Holland, Vasilis Friderikos, Mischa Dohler and Hamid Aghvami, ``Virtualized Authentication, Authorization and Accounting (V-AAA) in 5G Networks'', \emph{IEEE Conference on Standards for Communications and Networking (CSCN)}, September 2017.
		
		\bibitem{3gpp2017system}
		3GPP Technical Specification Group Services and System Aspects, \emph{3GPP TS 23.501 V1.5.0: System Architecture for the 5G System; Stage 2 (Release 15)}, November 2017.
		
		\bibitem{5gppp2016white}
		5G PPP Architecture Working Group, \emph{View on 5G Architecture}, White Paper, July 2016.
		
		\bibitem{han2017security}
		Bin Han, Stan Wong, Christian Mannweiler, Mischa Dohler and Hans D. Schotten, ``Security Trust Zone in 5G Networks'', \emph{24th International Conference on Telecommunications (ICT)}. IEEE, 2017.
		
		
		\bibitem{checko2015cloud}
		Aleksandra Checko, Henrik Lehrmann Christiansen, Ying Yan, Lara Scolari, Georgios Kardaras, Michael St\"ubert Berger and Lars Dittmann, ``Cloud RAN for Mobile Networks -- A Technology Overview'', \emph{IEEE Communications Surveys \& Tutorials} Vol. 17, Issue 1, pp. 405--426. IEEE, 2015.
		
		\bibitem{ge2016user}
		Xiaohu Ge, Junliang Ye, Yang Yang and Qiang Li, ``User Mobility Evaluation for 5G Small Cell Networks Based on Individual Mobility Model'', \emph{IEEE Journal on Selected Areas in Communications}, Vol. 34, Issue 3, pp. 528--541. IEEE, 2016.		
		\change{\bibitem{pukite1998markov}
		Paul Pukite and Jan Pukite, ``Markov Modeling for Reliability Analysis'', Wiley-IEEE Press, 1998.}
		\change{\bibitem{ye2018}
		Junliang Ye, Xiaohu Ge, Guoqiang Mao and Yi Zhong, ``5G Ultradense Networks With Nonuniform Distributed Users'',\emph{ IEEE Transactions on Vehicular Technology}, Vol. 67, Issue 3, pp. 2660--2670. IEEE, 2018.
		}
		
		
	\end{thebibliography}
\end{document}